# Observation of classical to quantum crossover in electron glass

Hideaki Murase[1], Shunto Arai[1], Takuro Sato[1,+], Kazuya Miyagawa[1], Hatsumi Mori[2], Tatsuo Hasegawa[1], and Kazushi Kanoda[1,*]

[1]*Department of Applied Physics, The University of Tokyo, Bunkyo-ku, Tokyo 113-8656, Japan.*

[2]*The Institute for Solid State Physics, The University of Tokyo, Kashiwa, Chiba 277-8581, Japan.*

[+]*Institute for Molecular Science, Myodaiji, Okazaki, 444-8585, Japan*

*Corresponding author. Email: kanoda@ap.t.u-tokyo.ac.jp

**Glass, a ubiquitous state of matter like a frozen liquid, is a seminal issue across fundamental and applied sciences and has long been investigated in the framework of classical mechanics[1,2]. A challenge in glass physics is the exploration of the quantum-mechanical behaviour of glass[3,4]. Experimentally, however, the real quantum manifestation of glass and the relationship between classical and quantum glass are totally unknown and remain to be observed in real systems. Here, we report the direct observation of classical-to-quantum evolution in the frustration-induced charge glass state exhibited by interacting electrons in organic materials[5,6]. We employ Raman spectroscopy to capture a snapshot of the charge density distribution of each molecule in a series of charge glasses formed on triangular lattices with**

**different geometrical frustrations. In less frustrated glass, the charge density profile exhibits a particle-like two-valued distribution; however, it becomes continuous and narrowed with increasing frustration, demonstrating the classical-to-quantum crossover. Moreover, the charge density distribution shows contrasting temperature evolution in classical and quantum glasses, enabling us to delineate energy landscapes with distinct features. The present result is the first to experimentally identify the quantum charge glass and show how it emerges from classical glass.**

In the science of glass, a new class of matter has emerged – electron glass (called charge glass to distinguish it from spin glass) formed by Coulomb interacting electrons on a triangular molecular lattice that fail to undergo Wigner crystallization owing to the lattice mismatch[5,7–9]. Charge glass qualitatively differs from the conventional glass formed by atoms, molecules, polymers, and colloids in two respects[1,2,10]. One is that charge glass forms on a regular lattice, which gives a new dimension to glass research, as it provides an opportunity to control the stage for the glass formation. The other is that the glass former has a quantum nature, which introduces the discipline of quantum physics to the glass physics thus far developed in the realm of classical physics. It is very intriguing to explore the unrevealed quantum nature of charge glasses with varying underlying lattice geometries; such experimental challenge will become possible with layered organic conductors, $\theta$-$(ET)_2X(SCN)_4$ (X=TlCo, RbZn and CsZn) hereafter referred to as $\theta$-X (Fig. 1a)[11].

In the conducting layers of $\theta$-X, the ET molecules form isosceles triangular lattices on which strongly interacting holes in a quarter-filled band suffer from strong charge frustration that is characterized by the anisotropy of intersite Coulomb energy, $V_p/V_c$ (Fig. 1b). The value of $V_p/V_c$ varies as 0.84, 0.87 and 0.91 in $\theta$-TlCo, $\theta$-RbZn and $\theta$-CsZn, respectively, according to molecular orbital calculations[12,13]. The degree of frustration plays a critical role in the nonequilibrium phase diagram of θ-X (Fig. 1c). At high temperatures, electrons in all the three systems are in the charge liquid (CL) state with moderate conductivity but largely exceeding the Mott-Ioffe-Regel limit[14], where charges fluctuate in space and time. In the least frustrated system, θ-TlCo, the CL undergoes a transition to charge order (CO) of insulating nature at 245 K[15]; however, when cooled extremely rapidly (>$10^3$ K/s), the CL freezes into charge glass (CG) instead of the CO transition[16–18]. The moderately frustrated θ-RbZn also undergoes a CO transition at 200 K during slow cooling (<1 K/min)[19–21], whereas when cooled quickly (>5 K/min), the CO is similarly suppressed and gives way to CG[5]. In the most frustrated system, θ-CsZn, the CL no longer falls into CO but instead freezes to CG, no matter how slowly it is cooled on a laboratory time scale[6]. Thus, the critical cooling speed to attain CG is highly dependent on the charge frustration; a more frustrated system is better able to form glass (Fig. 1c).

The CG in θ-X exhibits the hallmarks of conventional glasses, such as slow fluctuations, medium length-scale correlation, aging and crystallization[5,6,16,22,23]. Remarkably, however, the conductivity, which indicates the occurrence of charge transfer more than ten orders of magnitude faster than the CG dynamics, and the structureless NMR spectra, which are inconsistent with random charge configurations, imply the quantum or wave nature of CG. In the present work, aiming to elucidate the quantum-classical or wave-particle duality behind CG formation, we investigate the microscopic charge density profiles of CL/CG and its frustration dependence through Raman spectroscopy.

Raman spectroscopy is a powerful method of investigating the microscopic charge density profile in organic conductors[20,24–29]. One of the C=C stretching modes of an ET molecule, $\nu_2$, is charge-sensitive, and its peak wavenumber is empirically known to vary linearly with charge density on an ET molecule[20,30]. The time scale at which the Raman spectrum detects charge density is that of the molecular vibration (~45 THz), which is much faster than that of the glassy slow fluctuation (kHz or slower) and even exceeds the characteristic frequency of thermal fluctuation (~2-6 THz=80-300 K) by an order of magnitude (Fig. 1d). Hence, the Raman spectrum provides a snapshot of the molecule-to-molecule charge-density distribution. The Raman spectrum, $I(\nu)$, was converted to the charge-density distribution, $D(\rho)$, with the valence-dependent Raman tensor taken into consideration (see Methods and Supplementary Notes I and II for details).

Fig. 2a shows the $D(\rho)$ of the CO state of θ-TlCo and θ-RbZn at 100 K attained by slow cooling, at 1 K/min. Both systems exhibit two sharp peaks at $\rho$=0.12 and 0.86 corresponding to the absence and presence of a charge on ET, respectively (Fig. 2b), indicating particle-like charge localization; the slight deviations from $\rho$=0 and 1 are due to the bare extension of the wave function to adjacent sites. Figs. 2c, 2e and 2g show a comparison of the $D(\rho)$ of the CG states of θ-TlCo, θ-RbZn and θ-CsZn at 82 K, which illustrates the frustration dependence of $D(\rho)$. When cooled extremely rapidly using pulsed photo or joule heating (Supplementary Note III), the least frustrated system, θ-TlCo, shows a charge-density profile with double peaks akin to that for the CO state; however, the peaks are reduced and slightly broadened, and there appears a continuum in an extended range of $\rho$ (Fig. 2c). As discussed later, the temperature dependence of $D(\rho)$ suggests that the two-peak structure is not due to the remnant of the CO domains but an intrinsic feature of the CG of θ-TlCo. The remaining two-valued feature of $D(\rho)$ suggests that the CG remains particle-like but probably has positional disorder, which may cause the reduction of the peaks and the emergence of the continuum (Fig. 2d). In the more frustrated θ-RbZn system, the peaks are considerably suppressed, and the distribution is extended with a broad peak at approximately $\rho = 0.5$ (Fig. 2e). The most frustrated θ-CsZn no longer shows any structure but exhibits a sharp single peak at $\rho = 0.5$ (Fig. 2g).

These systematic results indicate that frustration narrows the inhomogeneous charge-density distribution. This is considered as a quantum effect. As the frustration is strengthened, an increasing number of different charge configurations that are degenerate become quantum mechanically mixed through the configuration interaction. This mixing extends the wave functions, and the charge-density disproportionation becomes less prominent. Note that even θ-CsZn, which has prominent quantum features, exhibits glass phenomena such as slow dynamics, middle-scale correlation and nonequilibrium aging[6], which dictates that θ-CsZn offers a clear substantiation of what should be called quantum glass. Intriguingly, the CG in θ-CsZn does not transition to the CO state no matter how slowly it is cooled on the laboratory time scale, possibly suggesting that the quantum nature stabilizes the glass state. Albeit in a different context from the present case, a numerical study suggests that quantum fluctuations can promote glass formation[4].

The temperature dependence of $D(\rho)$ on warming from a CG to CL gives insight into the topographical profile of the energy landscape and its frustration dependence. Figs. 3a-3c show the temperature evolution of $D(\rho)$ for θ-TlCo, θ-RbZn and θ-CsZn measured upon warming up after quenching to 82 K, a temperature at which CG is metastabilized (see Supplementary Note IV for all the measured Raman spectra and deduced $D(\rho)$). The temperature evolution of the width of $D(\rho)$ defined by the standard deviation is displayed in Fig. 3d; the data of θ-TlCo are lacking in 150-240 K where fast crystallization makes it difficult to secure enough time for acquiring Raman spectra of CG.

In the least frustrated θ-TlCo, the low-temperature double-peak structure is reduced in peak height and the intensity of the continuum increases upon warming from 82 K to 140 K (see also Supplementary Note VI). If the peaks come from the mixed domains of CO, the peak height increases at higher temperatures reached later after quenching, contrary to the observation. Thus, the double-peak structure and its temperature evolution are intrinsic to CG. The double-peak structure largely changes to a structureless shape at

250 K and is somewhat narrowed at 300 K. Although the spectral features appear distinct below 140 K and above 250 K, the temperature evolution of the distribution width suggests that the profiles of $D(\rho)$ observed below 140 K and above 250 K are parts of the continuous change of $D(\rho)$ on the way from CG to CL. The results indicate that the charge-density distribution in θ-TlCo becomes homogeneous with increasing temperature. The narrowing of the $D(\rho)$ profile with increasing temperature is also commonly observed in θ-RbZn (Figs. 3b and 3d). As mentioned above, the characteristic frequency of thermal fluctuations, $\sim k_B T/h$, is much lower than the Raman observation frequency; namely, the narrowing is not dynamic but results from the statistical ensemble of the thermally excited charge configurations. This result suggests that charge configurations in CG are particle-like at low energies and take on extended wave-like features at higher energies, leading to less disproportionation.

Remarkably, however, the most frustrated θ-CsZn exhibits a distinctive temperature evolution of $D(\rho)$, whose distribution width, considerably narrowed in CG at low temperatures, increases with temperature as opposed to the behaviour of θ-TlCo and θ-RbZn (Figs. 3c, 3d). Considering that $D(\rho)$ reflects the temperature-dependent ensemble of activated charge configurations, the behaviour of θ-CsZn suggests that higher energy states have more localized charge configurations unlike the other two systems, which illustrates the distinct energy dependence of the charge-density profile between the quantum and classical CG's. Interestingly, at the high temperature limit, the $D(\rho)$ values for the three systems appear to approach a common distribution form with $\Delta\rho \sim 0.15$. It is likely that, at increasing temperatures, intense lattice vibrations make the stroboscopic lattice pattern so disordered as to be insensitive to the difference in lattices among these three systems.

From the present results, we propose a conceptual topography of the energy landscape for classical to quantum charge glasses, as shown in Fig. 4, where the horizontal axis represents the charged-particle configurations on an anisotropic triangular lattice[31]. A large basin determines the global energy profile, where the thermal activation from a basin to adjacent ones may give classical slow dynamics responsible for the so-called α relaxation, which is suppressed in CG (refs. [32,33]). Within a large basin, there are numerous small basins corresponding to various local particle configurations, between which β relaxation may occur.

In the less frustrated case (Fig. 4a), charge configurations at low energies are less degenerate and less mixed so that the particle character is preserved in the glass state. At higher energies, the density of states of the small basins is larger so that the increased degeneracy promotes mixing between particle configurations, and the CG bears wave-like characteristics, explaining the narrowing of the $D(\rho)$ profile with increasing temperature. This result is reminiscent of the energy-dependent localization and delocalization profile in the density of states in the Anderson localization (Fig. 4a)[34]. In the strongly frustrated system (Fig. 4b), numerous configurations that are degenerate near the lowest energy are hybridized to form a new set of eigenstates with more extended wave functions, which are signified by a vanishingly small charge transport gap in θ-CsZn, compared with the gap, ~300 K, in θ-RbZn[35]. This energy topography explains the narrow width of $D(\rho)$ in CG at low temperatures, and the decrease in degeneracy at elevated energies, unlike the above classical case, explains the broadening of $D(\rho)$. Even in a system with a strong quantum nature, if basins remain in the landscape, the system retains the properties of classical glass along with its quantum nature. This hierarchical structure of the energy landscape provides a picture of quantum charge glass, as substantiated in θ-CsZn.

In the present work, we succeeded in directly observing the evolution from classical to quantum glasses. The charge-density profile, $D(\rho)$, probed by Raman spectroscopy captured how the charge glass formed by interacting electrons on triangular lattices takes on a quantum nature with varying geometrical frustration and temperature. The frustration dependence of $D(\rho)$ illustrates a route from classical to quantum glass, and the temperature variation of $D(\rho)$ provides insight into the topographical structure of the energy landscape, which is distinctive between classical and quantum glasses. We note that a more strongly frustrated system, θ-$I_3$, no longer behaves as CG but becomes a Fermi liquid at low temperatures[30], which can be regarded as quantum melting of CG due to the strong configuration interactions between large basins. All of the above observations point to a global scenario for the fate of glass formation by interacting quantum particles with increasing frustration; specifically, classical glass transitions to quantum glass and eventually undergoes quantum melt.

Reference

1. Lubchenko, V. & Wolynes, P. G. Theory of structural glasses and supercooled liquids. *Annu. Rev. Phys. Chem.* **58**, 235–266 (2007).

2. Debenedetti, P. G. & Stillinger, F. H. Supercooled liquids and the glass transition. *Nature* **410**, 259–267 (2001).

3. Müller, M., Strack, P. & Sachdev, S. Quantum charge glasses of itinerant fermions with cavity-mediated long-range interactions. *Phys. Rev. A - At. Mol. Opt. Phys.* **86**, 1–16 (2012).

4. Markland, T. E. *et al.* Quantum fluctuations can promote or inhibit glass formation. *Nat. Phys.* **7**, 134–137 (2011).

5. Kagawa, F. *et al.* Charge-cluster glass in an organic conductor. *Nat. Phys.* **9**, 419–422 (2013).

6. Sato, T. *et al.* Emergence of nonequilibrium charge dynamics in a charge-cluster glass. *Phys. Rev. B - Condens. Matter Mater. Phys.* **89**, 1–5 (2014).

7. Schmalian, J. & Wolynes, P. G. Stripe glasses: self-generated randomness in a uniformly frustrated system. *Phys. Rev. Lett.* **85**, 836–839 (2000).

8. Mahmoudian, S., Rademaker, L., Ralko, A., Fratini, S. & Dobrosavljević, V. Glassy Dynamics in Geometrically Frustrated Coulomb Liquids without Disorder. *Phys. Rev. Lett.* **115**, 1–5 (2015).

9. Yoshimi, K. & Maebashi, H. Coulomb frustrated phase separation in quasi-two-dimensional organic conductors on the verge of charge ordering. *J. Phys. Soc. Japan* **81**, 1–4 (2012).

10. Hunter, G. L. & Weeks, E. R. The physics of the colloidal glass transition. *Reports Prog. Phys.* **75**, (2012).

11. Mori, H., Tanaka, S. & Mori, T. Systematic study of the electronic state in θ-type BEDT-TTF organic conductors by changing the electronic correlation. *Phys. Rev. B* **57**, 12023–12029 (1998).

12. Mori, T. Non-stripe charge order in the θ-phase organic conductors. *J. Phys. Soc. Japan* **72**, 1469–1475 (2003).

13. Kondo, R. *et al.* Electrical and structural properties of θ-type BEDT-TTF organic conductors under uniaxial strain. *J. Phys. Soc. Japan* **75**, 1–7 (2006).

14. Sato, T. *et al.* Strange metal from a frustration-driven charge order instability. *Nat. Mater.* 1 (2019) doi:10.1038/s41563-019-0284-9.

15. Seo, H. Charge ordering in organic ET compounds. *J. Phys. Soc. Japan* **69**, 805–820 (2000).

16. Sato, T. *et al.* Systematic variations in the charge-glass-forming ability of geometrically frustrated θ-(BEDT-TTF)$_2$X organic conductors. *J. Phys. Soc. Japan* **83**, 1–4 (2014).

17. Miyagawa, K. *et al.* Charge Order and Poor Glass-forming Ability of an Anisotropic Triangular-lattice System, θ-(BEDT-TTF)$_2$TlCo(SCN)$_4$, Investigated by NMR. *J. Phys. Soc. Japan* **88**, 1–5 (2019).

18. Oike, H. *et al.* Phase-change memory function of correlated electrons in organic conductors. *Phys. Rev. B - Condens. Matter Mater. Phys.* **91**, 2–5 (2015).

19. Miyagawa, K., Kawamoto, A. & Kanoda, K. Charge ordering in a quasi-two-dimensional organic conductor. *Phys. Rev. B - Condens. Matter Mater. Phys.* **62**, R7679–R7682 (2000).

20. Yamamoto, K., Yakushi, K., Miyagawa, K., Kanoda, K. & Kawamoto, A. Charge ordering in θ-(BEDT−TTF)$_2$RbZn(SCN)$_4$ studied by vibrational spectroscopy. *Phys. Rev. B* **65**, 085110 (2002).

21. Watanabe, M., Noda, Y., Nogami, Y. & Mori, H. Investigation of X-ray diffuse scattering in θ-(BEDT-TTF)$_2$RbM'(SCN)$_4$. *Synth. Met.* **135–136**, 665–666 (2003).

22. Sato, T., Miyagawa, K. & Kanoda, K. Electronic crystal growth. *Science* **357**, 1378–1381 (2017).

23. Sasaki, S. *et al.* Crystallization and vitrification of electrons in a glass-forming charge liquid. *Science* **357**, 1381–1385 (2017).

24. Ouyang, J., Yakushi, K., Misaki, Y. & Tanaka, K. Raman spectroscopic evidence for the charge disproportionation in a quasi-two-dimensional organic conductor θ-(BDT-TTP)$_2$Cu(NCS)$_2$. *Phys. Rev. B - Condens. Matter Mater. Phys.* **63**, 054301 (2001).

25. Drozdova, O. *et al.* Raman spectroscopy as a method of determination of the charge on BO in its complexes. *Synth. Met.* **120**, 739–740 (2001).

26. Wang, H. H., Ferraro, J. R., Williams, J. M., Geiser, U. & Schlueter, J. A. Rapid Raman spectroscopic determination of the stoichiometry of microscopic quantities of BEDT-TTF-based organic conductors and superconductors. *J. Chem. Soc. Chem. Commun.* 1893–1894 (1994) doi:10.1039/C39940001893.

27. Wang, H. H., Kini, A. M. & Williams, J. M. Raman characterization of the BEDT-TTF(ClO$_4$)$_2$ salt. *Mol. Cryst. Liq. Cryst. Sci. Technol. Sect. A Mol. Cryst. Liq. Cryst.* **284**, 211–221 (1996).

28. Moldenhauer, J. *et al.* FT-IR absorption spectroscopy of BEDT-TTF radical salts: charge transfer and donor-anion interaction. *Synth. Met.* **60**, 31–38 (1993).

29. Yakushi, K. Infrared and Raman Studies of Charge Ordering in Organic Conductors, BEDT-TTF Salts with Quarter-Filled Bands. *Crystals* **2**, 1291–1346 (2012).

30. Yamamoto, T. *et al.* Examination of the charge-sensitive vibrational modes in bis(ethylenedithio)tetrathiafulvalene. *J. Phys. Chem. B* **109**, 15226–15235 (2005).

31. Stillinger, F. H. A topographic view of supercooled liquids and glass formation. *Science* **267**, 1935–1939 (1995).

32. Johari, G. P. Intrinsic mobility of molecular glasses. *J. Chem. Phys.* **58**, 1766–1770 (1973).

33. Rössler, E., Warschewske, U., Eiermann, P., Sokolov, A. P. & Quitmann, D. Indications for a change of transport mechanism in supercooled liquids and the dynamics close and below $T_g$. *J. Non. Cryst. Solids* **172**, 113–125 (1994).

34. Lee, P. A. & Ramakrishnan, T. V. Disordered electronic systems. *Rev. Mod. Phys.* **57**, 287–337 (1985).

35. Sato, T., Miyagawa, K., Tamura, M. & Kanoda, K. Anomalous 2D-Confined Electronic Transport in Layered Organic Charge-Glass Systems. *Phys. Rev. Lett.* **125**, 146601 (2020).

**Acknowledgments.** The authors thank A. Ikeda, F. Kagawa, M. Ogata, R. Furukawa, T. Kato and K. Yoshimi for their fruitful discussion and the Cryogenic Research Center at the University of Tokyo for supporting low-temperature experiments. This work was supported by Japan Society for the Promotion of Science (JSPS) under Grant Numbers 18H05225, 18H05516, 19H01846, 20K20894, 20KK0060, 19H02579, 21H05234 and 21K18597, and the Mitsubishi Foundation under Grant Number 202110014. S.A. also thanks to the support from Murata Science Foundation.

**Author contributions.** K.M., T.S, and H.M. prepared samples. H.M. performed experiments and analyzed as well as interpreted data with the help of S.A., T.H., K.M.

and K.K. K.K designed the project. H.M. and K.K. wrote the manuscript with the input from all authors.

**Competing financial interests.** The authors declare no competing financial interests.

**Additional information**

    **Supplementary information** is available for this paper at ***.

    **Correspondence and requests for materials** should be addressed to K.K.

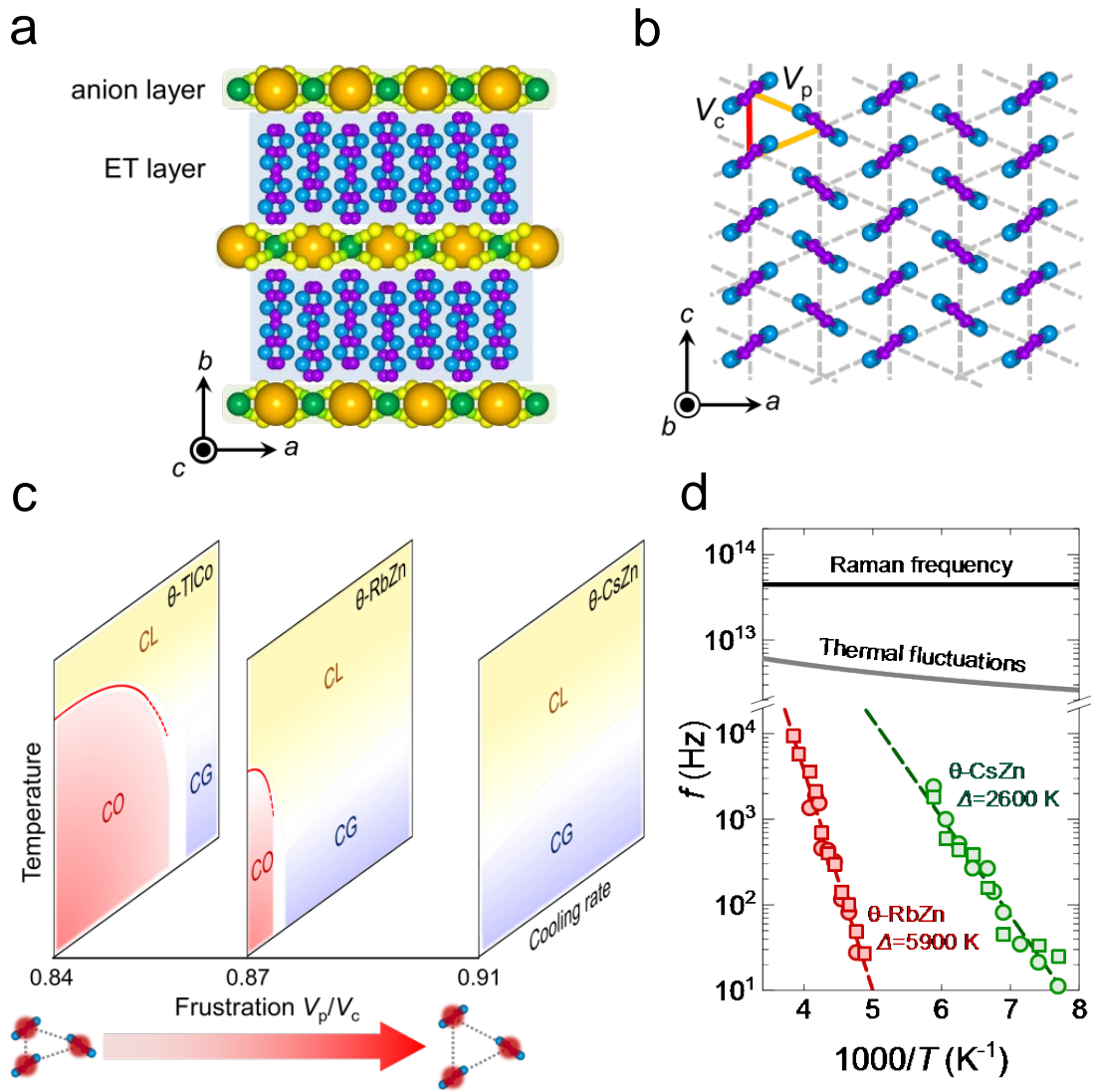

**Figure 1 | Crystal structure and phase diagram of θ-X.**
**a**, Layered crystal structure of θ-TlCo, θ-RbZn and θ-CsZn. **b**, ET layer viewed from the b-axis. The ET molecules form an isosceles triangular lattice characterized by two intersite coulombic repulsive energies $V_p$ and $V_c$. **c**, Schematic phase diagram in the parameter space of temperature, cooling rate and frustration for θ-TlCo, θ-RbZn and θ-CsZn. CL, CG and CO represent charge liquid, charge glass and charge order, respectively. **d**, Raman observation frequency, thermal-fluctuation frequency and charge-density-fluctuation frequencies in the charge liquids in θ-RbZn and θ-CsZn reproduced from Refs. 5 and 6.

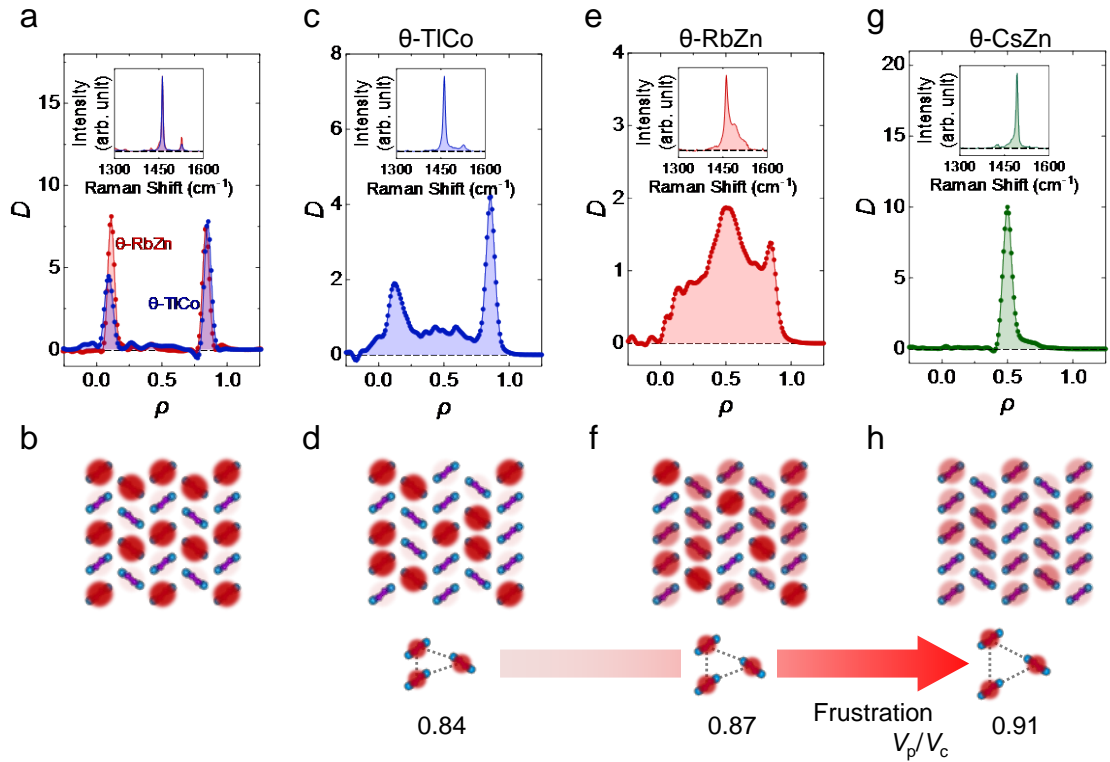

**Figure 2 | Frustration dependence of the charge density profile deduced from Raman spectra.**

Charge density distributions, $D(\rho)$, and schematic charge configurations in the CO phases of θ-TlCo and θ-RbZn at 100 K (a, b) and in the CG phases of θ-TlCo (c, d), θ-RbZn (e, f) and θ-CsZn (g, h) at 82 K are shown with the schematic representation of the anisotropy of the intersite Coulomb energies $V_p$ and $V_c$. Insets in (a, c, e, g) display Raman spectra, from which $D(\rho)$ is obtained.

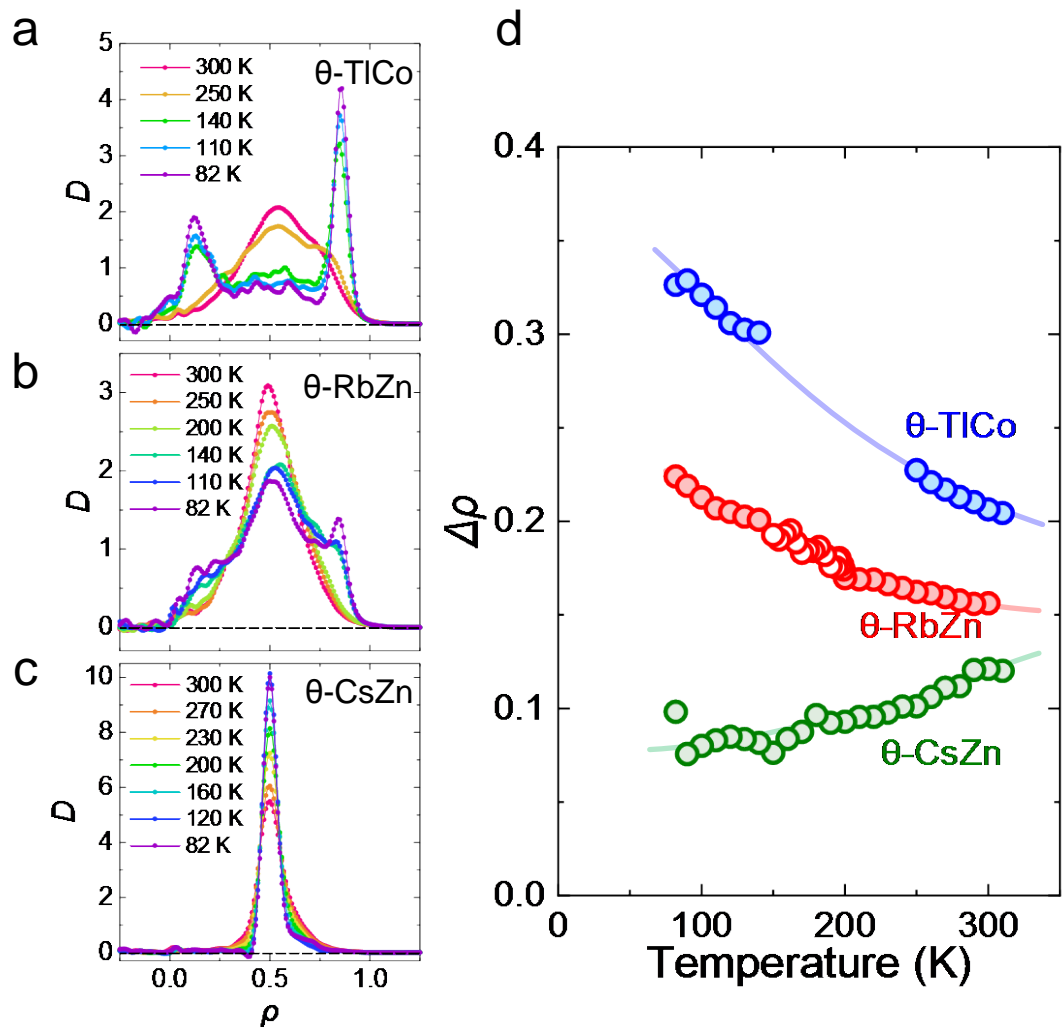

**Figure 3 | Temperature dependence of the charge density distributions derived from Raman spectra.**

**a**-**c**, Charge density distributions in θ-TlCo (a), θ-RbZn (b) and θ-CsZn (c) at various temperatures. **d**, Temperature dependences of the width of $D(\rho)$ for θ-TlCo, θ-RbZn and θ-CsZn. The width, $\Delta\rho$, is defined by the standard deviation of $D(\rho)$. The bold lines are visual guides. (See Supplementary Note II for detailed Raman spectra and $D(\rho)$ for the analysis). The measurements were performed in the sequence of raising the temperature. However, in the range of 150-200 K for θ-RbZn, where fast crystallization impeded the sequential measurements, each measurement was performed every time the sample was heated up to 210 K and rapidly cooled to the target temperature (open red circles in **d**) (Supplementary Note V).

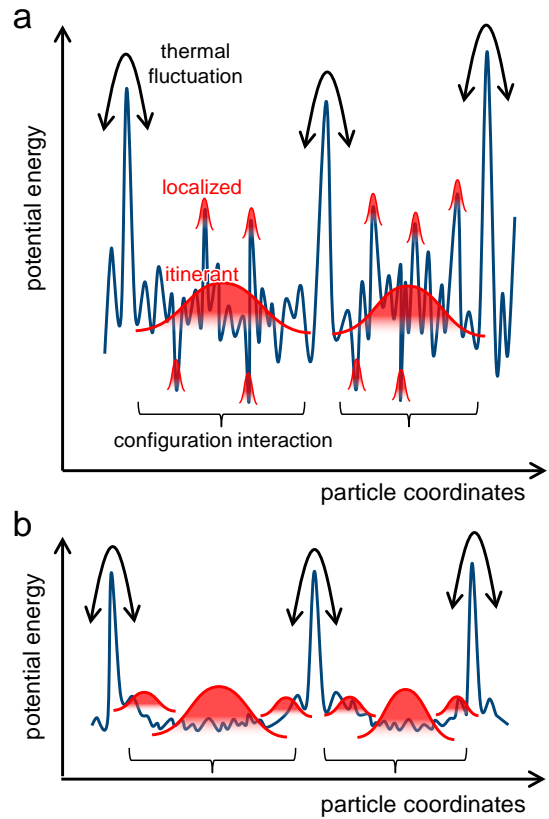

**Figure 4 | Energy landscape of charge glass.**
**a**, **b**, Energy landscape in less (a) and strongly (b) frustrated cases. The potential energy is plotted against particle coordinates; a point in the horizontal axis represents a particle-like charge configuration of the entire electrons on the lattice in an occupied/unoccupied manner. The real configuration of quantum mechanical particles is determined by the configuration interaction between the original configurations, which results in new quantum configurations depicted by red mountain shapes, each of which expresses the distribution of the weight of configurations that participate in the quantum configuration. Less degenerate particle configurations have deep energy minima or sharp energy maxima with poor configuration mixing, and thus, the particle-like nature is maintained at low energies (a). In the more degenerate case, strong configuration interactions that occur even at the lowest energies of the landscape cause the mixing of a large number of configurations to enhance the wave-like nature with less and continuous charge disproportionation (b). The configuration phase space is separated by large potential barriers to subspaces with no configuration interaction but thermal activation of the so-called α relaxation between them.

**Methods**

**Raman spectroscopy.** We employed Raman spectroscopy to probe the charge density profile of each molecule in the BEDT-TTF organic conductors[24]. One of the C=C stretching modes, $\nu_2$, which is particularly charge sensitive[25–28], was investigated to determine the charge density distribution, $D(\rho)$.

We used $^{13}$C-enriched single crystals of θ-TlCo, θ-RbZn and θ-CsZn, where the central double-bonded carbons in the BEDT-TTF molecule are enriched to $^{13}$C isotopes to 99% concentration, because the C=C stretching modes under investigation are well separated from other nearly degenerate molecular vibrations by the site-selective isotope substitution[20]. The typical crystal size was approximately $1 \times 0.1 \times 0.1$ mm$^3$. The crystals were mounted on copper substrates and then loaded on a cooling stage (Linkam 10002L), which had a glass window for optical measurements. The cooling rate was 1 K/min or slower except for the case of realizing CG of θ-TlCo and θ-RbZn. To metastabilize CG in θ-RbZn, we cooled it at a rate of 30 K/min. (See below for θ-TlCo.) The Raman spectra of CG in θ-TlCo and θ-RbZn were observed upon warming with a Renishaw inVia Raman spectroscopy system. Excitation was provided from a 532 nm laser focused through a microscope equipped with a Leica N Plan L50x objective lens. The laser power was 45 mW. To prevent laser heating, the line focus technique was used. The scattered light in the backscattering geometry was split by a diffraction grating of 1800 g/mm and detected by a charge coupling device. The excitation light polarized parallel to the a-axis was irradiated in a direction normal to the (010) surface, and the scattered light polarized parallel to the a-axis was collected. The investigated temperature range was 82-310 K. The accumulation time was 2 hours.

**Rapid cooling of θ-TlCo.** Because an extremely rapid cooling rate is needed to form CG in θ-TlCo, we utilized an electrical current pulse[18]. An electrical current pulse with 40 V between the terminals was applied to the sample for 1 s at 130 K so that a part of the sample was heated up to above the charge ordering temperature and rapidly cooled back to 130 K to be frozen to CG. To optimize the voltage applied to the sample, a load resistor of 30 kΩ was placed in series with the sample. The generation of CG by this heating and cooling process was confirmed by Raman spectroscopy.

**Analysis of Raman spectra.** The frequency of the charge-sensitive $\nu_2$ mode, $\nu$, is known to shift linearly with the charge density $\rho$ on the ET molecule in the form of $\nu$=1537 cm$^{-1}$-90$\rho$. The coefficient of 90 cm$^{-1}$/$e$ is different from the value observed in previous research[29,30], 120 cm$^{-1}$/$e$, because of the effect of $^{13}$C substitution. When $\rho$ is distributed, the Raman spectrum, $I(\nu)$, is given as the form of convolution: $I(\nu) = \int A(\rho)D(\rho)L(\nu - \nu'(\rho))d\rho$, where $A(\rho)$ is the square of the $\rho$-dependent Raman tensor, $D(\rho)$ is the charge density distribution, and $L(\nu)$ is the Lorentzian function expressing the natural broadening, whose width was determined to be 4 cm$^{-1}$ with reference to the Raman spectrum of the CO phase in θ-RbZn. The $\rho$ dependence of the Raman tensor is possibly related to the electron-molecular vibration coupling and is empirically assumed by the method described in the Supplementary Note I. Then, $D(\rho)$ is calculated as $D = A^{-1}\mathcal{F}^{-1}[\mathcal{F}[I]/\mathcal{F}[L]]$, where $\mathcal{F}$ expresses the operation of the Fourier transformation.

# SUPPLEMENTARY INFORATION

**Note I. Charge-density dependence of the squared Raman tensor, $A(\rho)$**

It is experimentally known that the square of the Raman tensor, $A$, which is necessary for bringing out the charge density distribution $D(\rho)$, depends on the charge density, $\rho$. Since $A(\rho)$ is proportional to the $\rho$-dependent Raman signal intensity, we determined it empirically using the experimental Raman spectral intensities for the four charge-order materials, $\theta$-TlCo, $\theta$-RbZn, $\alpha$-(ET)$_2$I$_3$ and $\kappa$-(ET)$_2$Hg(SCN)$_2$Cl. The value of $\kappa$-(ET)$_2$Hg(SCN)$_2$Cl was determined by referring to the literature[1]. As an appropriate functional form of $A(\rho)$, we adopt the form of $\exp(\rho/\rho_0)+a$ (Fig. S1a) to reproduce these data. Because the ratio of the spectral peak for the charge-rich molecule to that for the charge-poor molecule is experimentally known, we fitted $A(1-\rho_P)/A(\rho_P)$ to the experimental values (Figs. S1b). The fitting yields a=1192 and $\rho_0$=0.0991.

**Note II. Examples of analysis of the Raman spectra to obtain $D(\rho)$**

As examples of the analysis of Raman spectra of the $\nu_2$ mode, we show the spectra of $\theta$-RbZn in the CO phase at 100 K and in the CL phase at 300 K and the derived $D(\rho)$'s in Fig. S2. The Raman spectrum of the CO exhibits a double-peak structure (Fig. S2a). The two peaks at the low and high wavenumbers come from the charge-rich and charge-poor molecules, respectively. The difference of the peak height is due to the above discussed charge-density dependence of the Raman tensor. The derived $D(\rho)$ reasonably exhibits a double-peak structure (Fig. S2b). The Raman spectrum of CL exhibits a broad peak coming from inhomogeneous charge density (Fig. S2c) and correspondingly the derived $D(\rho)$ shows a broad distribution of $\rho$ with a peak around $\rho$ = 0.5 (Fig. S2d).

**Note III. Rapid cooling of θ-TlCo by current and laser pulses**

Extremely rapid cooling is required for realizing the CG in θ-TlCo. For this purpose, we employed two methods, which are to apply the current and laser pulses, where the cooling at the rate of ~$10^3$ K/s was attained.

A current pulse was applied through the electrodes of gold wires pasted onto the sample (Fig. S3a) while the equilibrium sample temperature is maintained to be 130 K. In the current pulse method, we chose this temperature as a trade-off between higher temperatures suitable for the Joule heating, which is not effective in an extremely large resistive state, and lower temperatures required for avoiding charge ordering. After the rapid cooling to 130 K attained by switching off the current pulse, the sample was cooled to a further lower temperature, 90 K, at which the possible nucleation of CO is more strongly prohibited. The real-space Raman image taken at this temperature in the case of the current pulse is shown in Fig. S3b, which displays the integral intensity of the Raman spectra between 1460 cm$^{-1}$ and 1463 cm$^{-1}$ in a range of colors and the Raman spectra in several spots in the sample. In the upper half region with red colored, the Raman spectra of CO were observed, whereas a somewhat different Raman spectrum was observed in the lower half region with yellow-green color (Fig. S3c). To verify that the new phase was generated by the rapid cooling, the sample was heated above the transition temperature of CO and cooled slowly to 90 K again. As a result, the Raman spectrum of CO were observed in the whole region (Fig. S3d). Because the new phase in the yellow-green region was reversibly generated by rapid cooling, it proves to be CG. By this technique, the CG phase of θ-TlCo was realized and the Raman spectra in the main text were observed.

The nearly the same result was obtained when we used the lase pulse. At 90 K, the laser of 2.3 W in power was irradiated for 0.5 s. To avoid overheating a local region, the laser was focused 50 μm away from the sample surface. Fig. S4a shows the contour plot of the integral intensity of the Raman spectra between 1460 cm$^{-1}$ and 1463 cm$^{-1}$ around the laser spot. In the outside of the spot (red region), the Raman spectra of CO were observed, whereas in the spot (yellow region), the Raman spectra of CG were observed. We also confirmed the reversibility of the CG (Fig. S4b). Fig. S4c shows the excellent agreement between the Raman spectra of CG realized by the laser and current pulse methods.

**Note IV. Raman spectra and charge density distributions**

In the main text in Figs. 2 and 3, we show the typical data of $D(\rho)$'s. All of the observed Raman spectra and corresponding $D(\rho)$'s are shown in Figs. S5, S6 and S7.

**Note V. Charge density distribution of θ-RbZn at temperatures of 150-200 K**

On cooling below $T_{CO}$ (200 K in θ-RbZn), the CL can be supercooled or frozen to CG and, the CL or CG evolvee to CO in the incubation time of ~4×10$^2$-10$^4$ s in θ-RbZn. Although the Raman spectra were measured at elevated temperatures in a sequentially warming process, we could not retain this measurement sequence in the temperature range of 150-200 K, where the relatively short incubation time inevitably allows for the nucleation of CO. Thus, in each measurement in this temperature range, we acquired the Raman spectrum for a short time before the nucleation of CO every time the sample is warmed up to 210 K (above $T_{CO}$) and rapidly cooled at 30 K/min to the target temperature from 210 K. To obtain intense Raman spectra in a short time, the spectra in this temperature range were acquired under particular conditions. One is the laser power, which was set to the high value of 225 mW. Although this value is five times larger than in the other measurements, we confirmed that the laser heating is negligible (< 1 K). Another is that a polarizing plate was not used. This means that not only the aa component but also ac component of the Raman tensor enter into the detection. Although the ac component may have a different $\rho$ dependence, the effect of the ac component proves to be negligible since the $\Delta\rho$ values in 150-200 K smoothly connect to those below 150K and above 200 K, as seen in in Fig. 3d. This is possibly because the ac component is weak enough in intensity or has the same $\rho$ dependence as the aa component. Figure S8 shows the charge density distributions $D(\rho)$ deduced from the Raman spectra measured for the limited times at 150, 158, 170, 182 and190 K. The resultant $D(\rho)$ at these temperatures, although somewat noisy, smoothly connect the behaviours below 150K and above 200 K, as seen in Fig. 3d.

**Note VI. Analysis of the temperature evolution of $D(\rho)$**

In the low-temperature GC states of θ-TlCo and θ-RbZn, the charge density distribution appears to consist of a two-valued component and a continuous one as shown in Fig. 3. As the CO state shows the similar spectral shape, one may suspect that the spectra in the CG contains CO domains. As we describe in the main text, that case is ruled out because the two-peak component decreases with increasing temperature; if the component came from the CO domains, it would grow on warming since the temperature-dependent measurements were performed in the warming process (temperature was fixed for each measurement). To characterize this temperature profile quantitatively, we fitted the spectra by the sum of Gaussians and obtained the volume fraction of the two-peak component. As an example, the fit of $D(\rho)$ of CG in θ-TlCo at 140 K is shown in Fig. S9a. The temperature dependences of the volume fraction of the two-peak component in θ-TlCo and θ-RbZn are shown in Fig. S9b, indicating that it decreases as the temperature increases in both systems and quantitatively confirming the conclusion above.

**Reference**
1. Hassan, N. M. *et al.* Melting of charge order in the low-temperature state of an electronic ferroelectric-like system. *npj Quantum Mater.* **5**, 1–6 (2020).

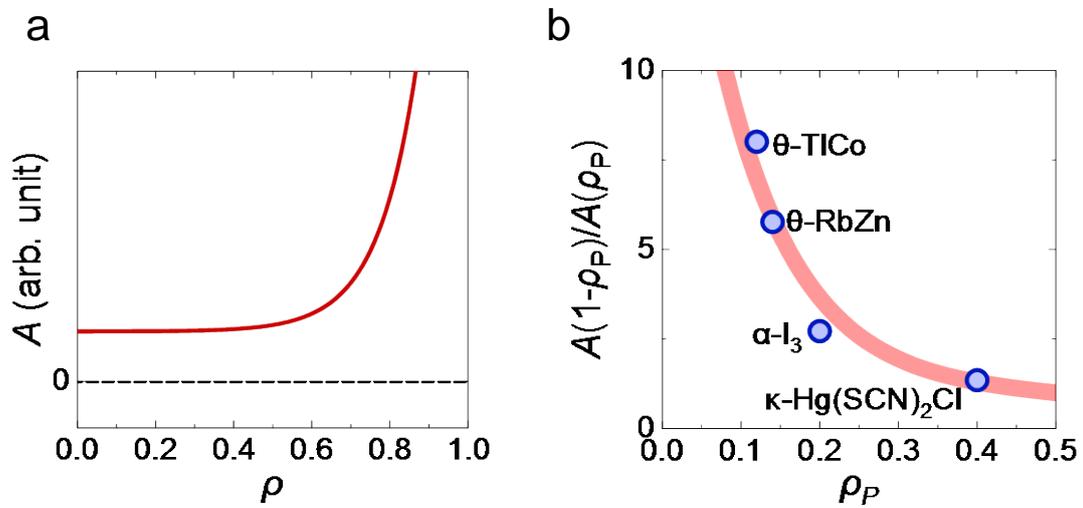

Fig. S1 Charge density dependence of the squared Raman tensor.
a, Charge density dependence of the square of the Raman tensor. b, Charge density dependence of the ratio of the Raman spectral peak for the charge-rich molecule to that for the charge-poor molecule. The experimental values of charge order materials for θ-TlCo, θ-RbZn, α-(ET)$_2$I$_3$ (α-I$_3$ in b) and κ-(ET)$_2$Hg(SCN)$_2$Cl are indicated by circles and the bold curve is the fit. The value of κ-(ET)$_2$Hg(SCN)$_2$Cl was determined with reference to the literature[1].

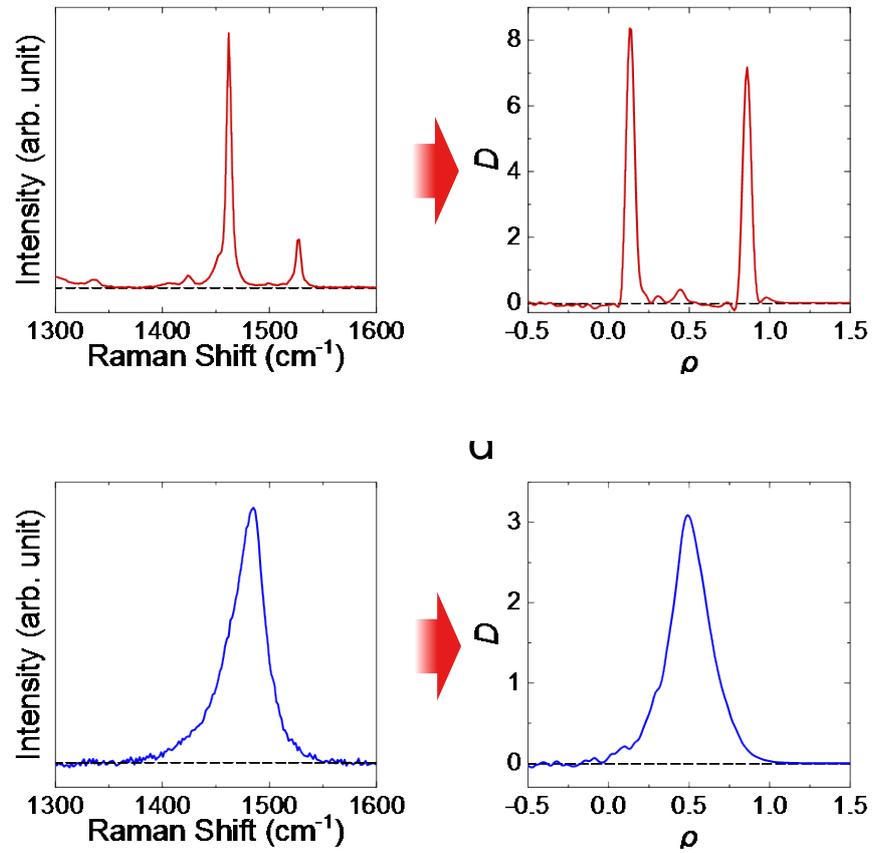

Fig. S2 Examples of the transformation of the Raman spectra to $D(\rho)$.
a,b, Raman spectrum of the CO phase in θ-RbZn at 100 K (a) and the derived $D(\rho)$ (b).
c,d, Raman spectrum of the CL phase in θ-RbZn at 300 K (c) and the derived $D(\rho)$ (d).

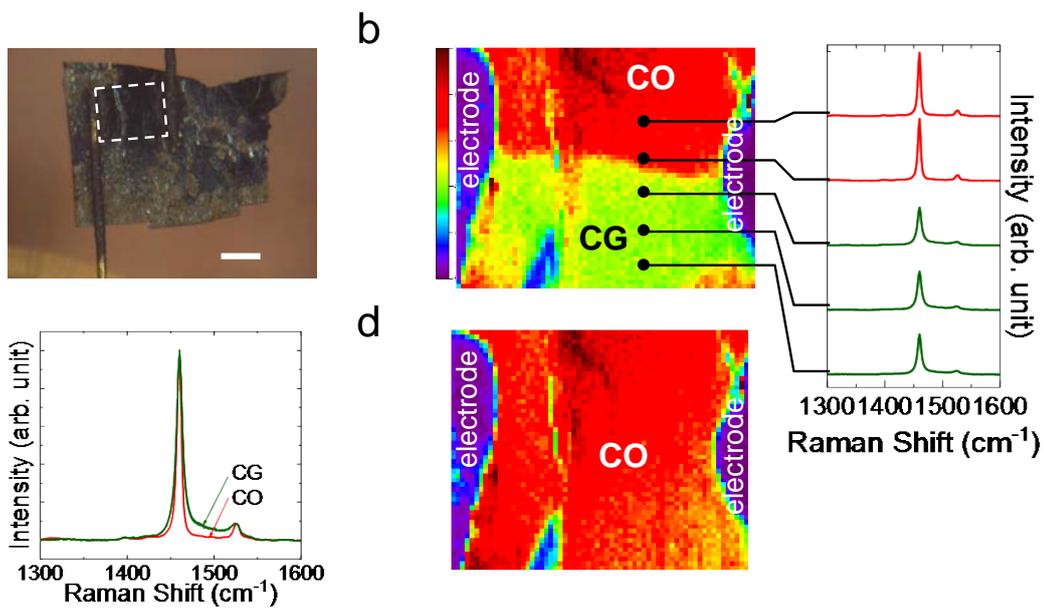

Fig. S3 Raman imaging and spectra of θ-TlCo after the rapid cooling with a current pulse.

a, Photograph of the sample with electrodes of gold wires. The square represented by the white dashed line is the region where Raman imaging was taken as shown in (b) and (d). b, Contour plot of the integral intensity of Raman spectrum after rapid cooling and the Raman spectra at several spots. c, Comparison of the Raman spectra shown in (b). d, Contour plot of the integral intensity of Raman spectrum after slow cooling. Refer to the supplementary text for the detailed experimental conditions.

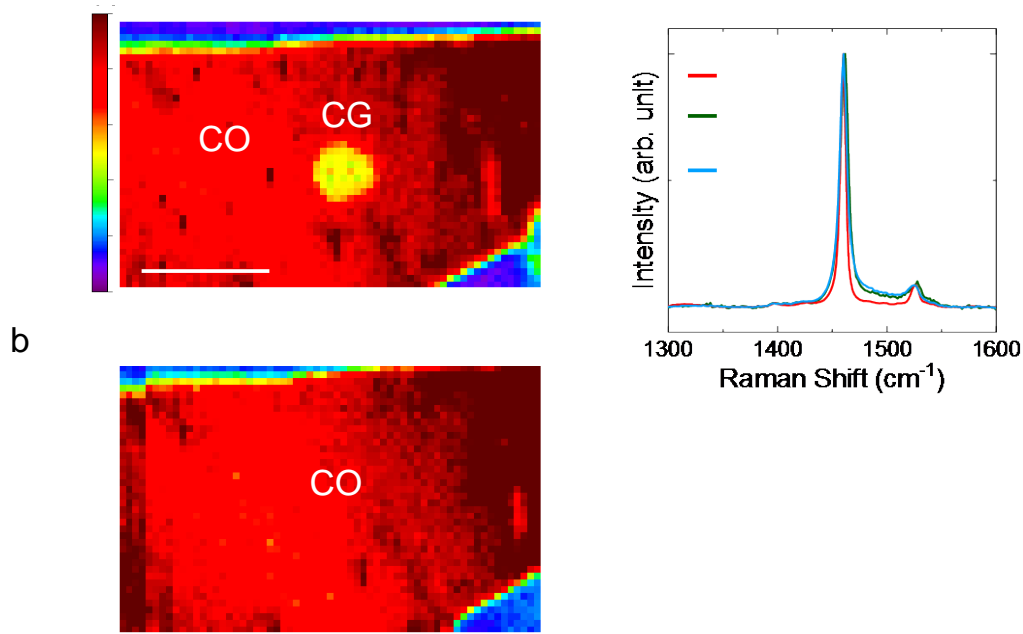

Fig. S4 Raman imaging and spectra of θ-TlCo after the rapid cooling with a laser pulse. a, Contour plot of the integral intensity of Raman spectra between 1460 cm$^{-1}$ and 1463 cm$^{-1}$ at 90 K after rapid cooling by the laser pulse irradiation. In the red region, the Raman spectra of CO were observed. A laser pulse was irradiated at the center of the yellow circle, where the Raman spectra of CG were observed. The blue region is the outside of the sample. b, Contour plot of the integral intensity. After taking the imaging of (a), the sample was heated above the transition temperature and cooled to 90 K again, then this imaging was taken. c, Comparison of the Raman spectra of CO, CG generated by the laser-pulse rapid cooling and CG generated by the current-pulse rapid cooling. The latter two spectra of CGs coincide well with each other.

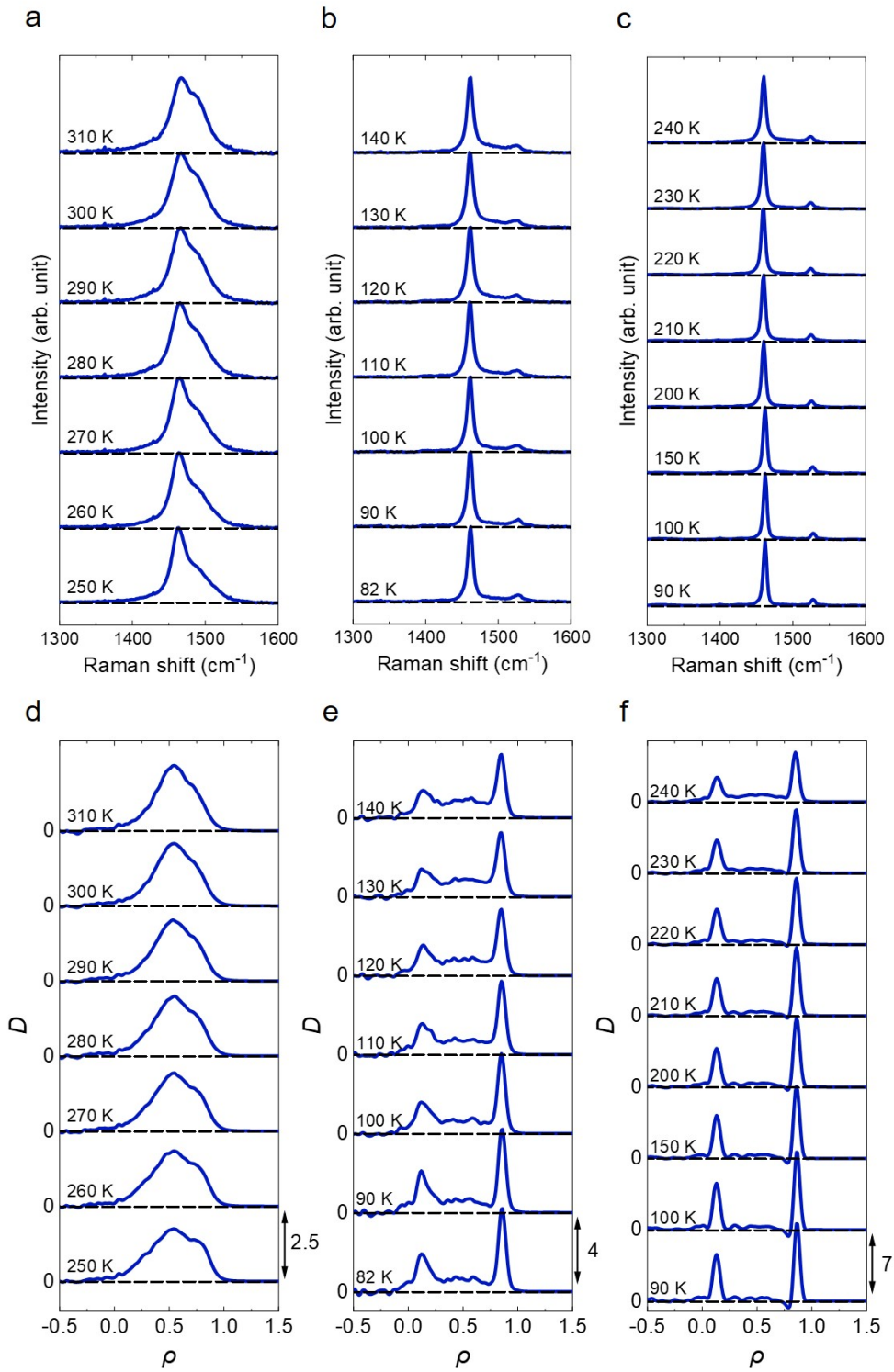

Fig. S5 Raman spectra and charge density distributions in θ-TlCo.
a-c, Raman spectra of θ-TlCo in CL (a), CG (b) and CO (c). d-f, Charge density distributions of θ-TlCo in CL (d), CG (e) and CO (f).

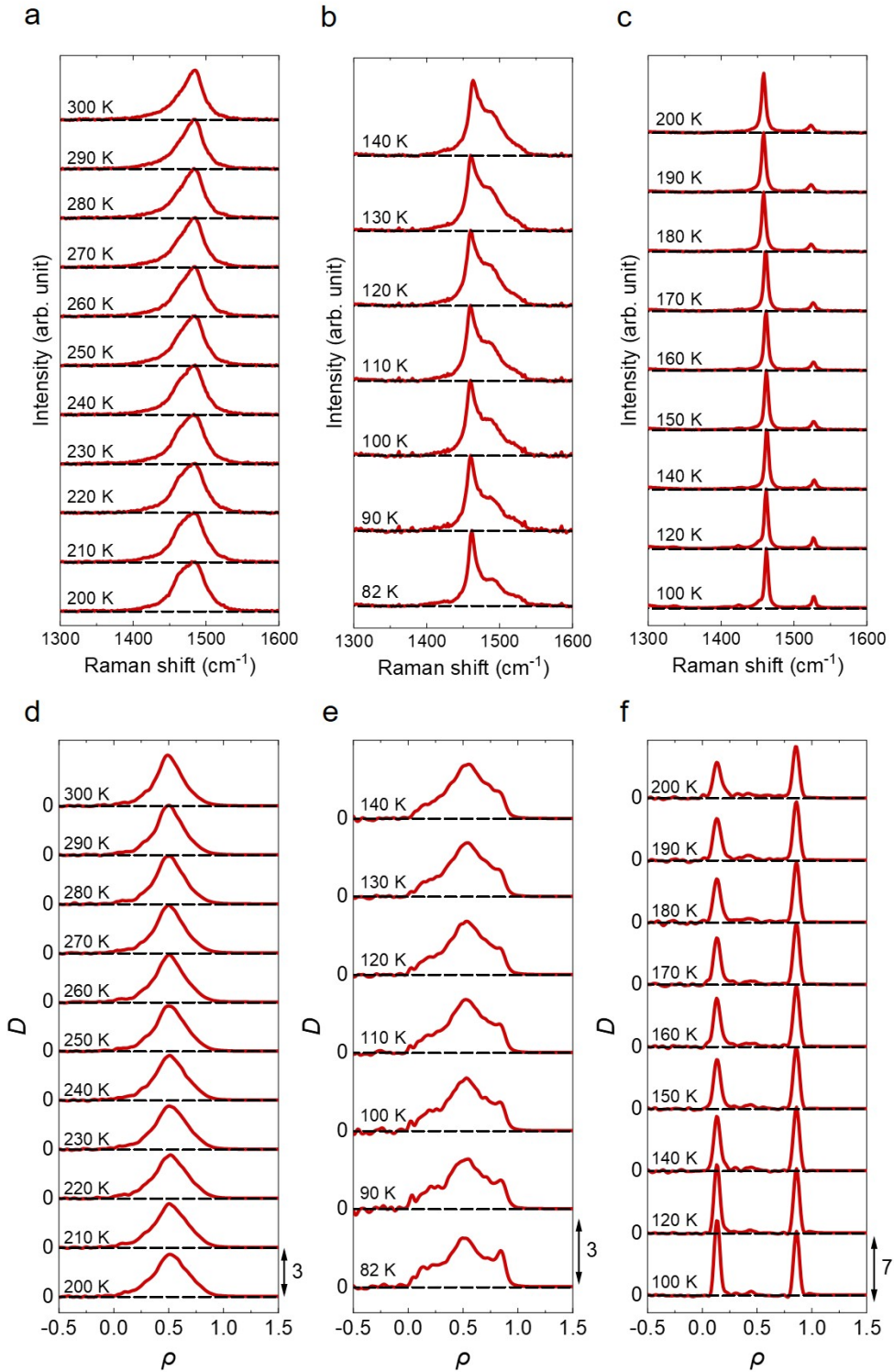

Fig. S6 Raman spectra and charge density distributions of θ-RbZn.
a-c, Raman spectra of θ-RbZn in CL (a), CG (b) and CO (c). d-f, Charge density distributions of θ-RbZn in CL (d), CG (e) and CO (f).

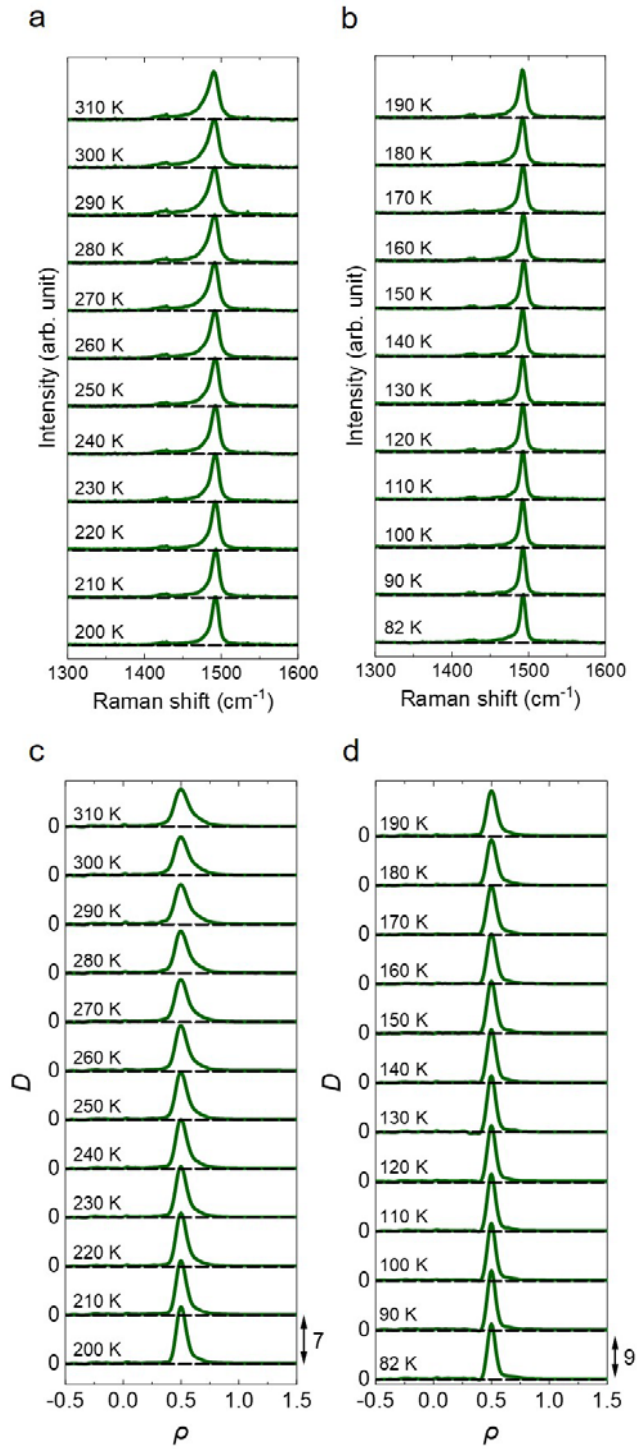

Fig. S7 Raman spectra and charge density distributions of θ-CsZn.
a,b, Raman spectra of θ-CsZn at high temperatures (a) and low temperatures (b). c,d, Charge density distributions of θ-CsZn at high temperatures (c) and low temperatures (d).

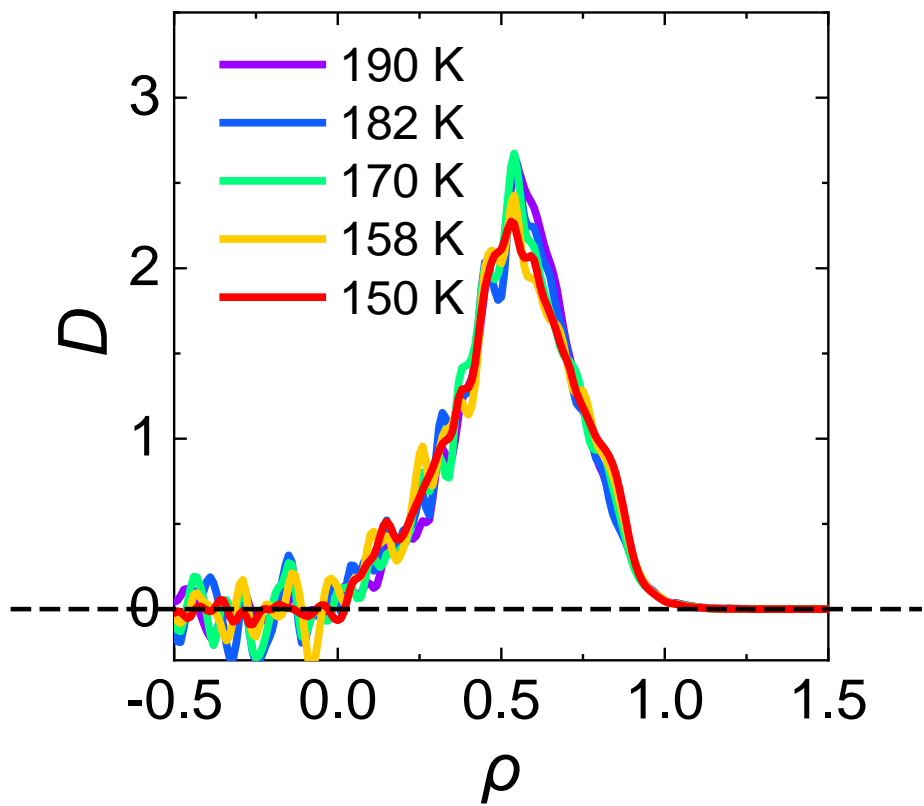

Fig. S8 Charge density distributions, $D(\rho)$, of the supercooled CL and CG in θ-RbZn in the temperature range of 150-200 K. In this temperature range, the spectra need to be acquired within the limited incubation time before the nucleation of CO.

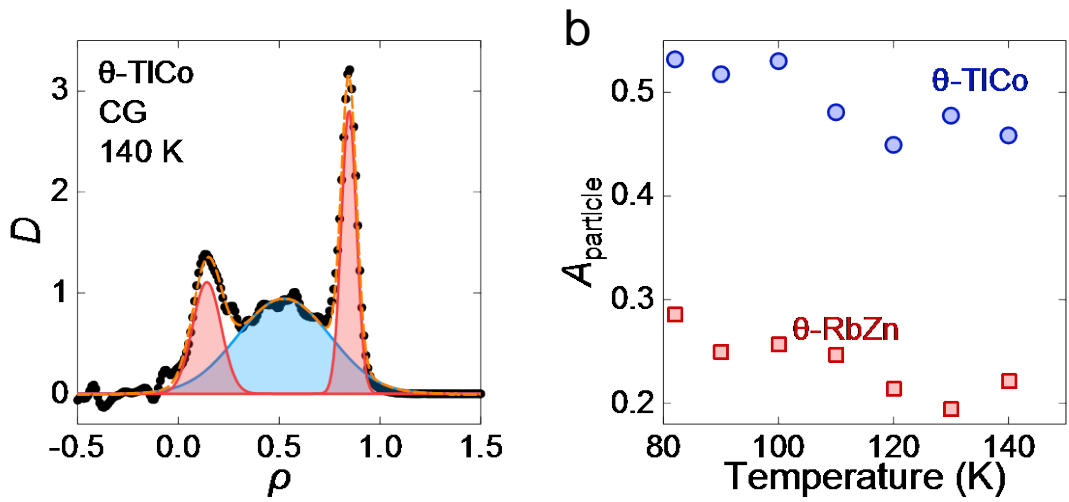

Fig. S9 Analysis of the charge density distribution. a, Fitting of the charge density distribution of CG in θ-TlCo at 140 K. The black points indicate the experimental data. The red and blue lines represent the fitting curves of the two-valued and broad peak components, respectively, and their sum is drawn by the orange dashed line. b, Temperature dependence of the two-peak component in θ-TlCo (blue circles) and θ-RbZn (red squares).